\documentclass[aps,preprint,tightenlines,floats,
groupedaddress,showpacs]{revtex4}
\usepackage{amssymb} 
\usepackage{hyperref} 
\usepackage{graphicx}
\usepackage{subfigure}

\begin{document}

\date{July 29, 2021}

\title{Clean energy from dark matter?}

\author{P. Sikivie}

\affiliation{Department of Physics, University of Florida, 
Gainesville, FL 32611, USA}

\begin{abstract}

A contribution to Frank Wilczek's 70th birthday's festschrift, 
this brief note considers how much power can be extracted from 
dark matter.

\end{abstract}

\maketitle

Frank Wilczek achieved great things as a theoretical 
physicist: asymptotic freedom of quantum chromodynamics, 
the axion as a consequence of the Peccei-Quinn solution 
to the Strong CP Problem, the proposal that the axion is 
the constituent particle of dark matter, anyonic quasi-particles 
in 2-dimensional condensed matter systems, and more yet.
Underlying these successes, I believe, is a view of physics 
as one large topic without the common place, but largely 
artificial, divisions into particle physics, cosmology 
and astrophysics, condensed matter physics, theory and 
experiment.  This view encourages one to explore outside 
one's own box.  Frank emboldened us when he said: ``If 
you are not making mistakes, you are working on things 
that are too easy."  So, his 70th birthday festschrift may 
be a good place to consider whether a significant amount 
of power can be extracted from dark matter.

To start off on a hopeful note, consider that the amount 
of energy stored in dark matter is truly enormous.  Our 
cosmic neighborhood, defined as a (kpc)$^3$ volume 
surrounding the Sun, contains on the order of 10$^{35}$ 
TeraWatt$\cdot$year in rest mass energy.  Whether it can 
be tapped as an energy source on Earth may be a laudable 
question.  It is certainly quixotic for two unrelated
reasons: 
\begin{itemize}
\item The density of dark matter on Earth is very small
\item The dark matter is very weakly coupled to ordinary 
matter.  
\end{itemize}
I'll argue below that if the dark matter is axions, or 
axion-like particles, the first difficulty largely 
disappears.  Unfortunately, the second difficulty 
remains for the time being.

The density of dark matter on Earth is commonly stated
to be $0.5 \times 10^{-24}$ gr/cc.  It should be kept 
in mind however that this number, derived from observation, 
signifies the average dark matter density on a length 
scale of order kpc ($\simeq 3 \times 10^{21}$ cm). 
Any dark matter detector on Earth is much smaller than 
that, by approximately 19 orders of magnitude.  The dark 
matter density on Earth may be very different from its 
kpc scale average, for example if the dark matter is 
clumpy.  In fact there is a straightforward reason why 
the dark matter density on Earth may be much larger than 
the above average value: dark matter particles form sharp 
caustics because they are collisionless and have very small 
primordial velocity dispersion.  At the caustics the dark 
matter density is much larger than average.

For the purpose of making estimates below, we will assume 
1) that dark matter has been detected on Earth and 2) 
that its density has been found to be the standard average 
value.  Let us keep in mind however that the density could 
be much larger.  The dark matter particles move with speeds 
of order 300 km/s since they form an extended halo around 
the Milky Way.  The rest mass energy flux density implied 
by the average density and typical velocity is 13 W/m$^2$.  
For comparison the solar energy flux density averaged over 
the sunlit half of the Earth is 680 W/m$^2$.

The most popular candidate for dark matter has been the 
WIMP, an acronym for Weakly Interacting Massive Particle.  
WIMP detectors on Earth are instrumented to detect collisions 
of WIMPs with nuclei in the bulk of the detector.  In each 
collision, the amount of energy deposited is at most of 
order the kinetic energy of the WIMP, which is 10$^{-6}$ 
times its rest mass energy. The largest average amount 
of kinetic energy is obtained when the nuclear target 
mass is of order the WIMP mass.  If all WIMPs going 
through a detector were to collide with nuclei in it, 
the amount of energy obtained would be of order 
$1.3 \times 10^{-5}$ W/m$^2$ times the surface area
of the detector.  But of course only a tiny fraction 
of WIMPs collide with nuclei.  Present upper limits 
on their scattering cross-section, of order 
$10^{-45}$ cm$^2$, imply that the scattering length 
of WIMPs through material bodies is of order kpc.  The 
notion of extracting significant power from WIMPs after 
they have been discovered to be the dark matter is 
truly a non-starter.

The situation is different and more tantalizing with 
the axion, the other leading dark matter candidate,
whose popularity has been rising lately \cite{CDEM}.
First, axions can be made to convert to photons or 
other forms of energy so that  they yield all 
their energy, their rest mass energy as well their
kinetic energy.  Second, dark matter axions are a 
highly degenerate Bose gas and therefore behave 
in first approximation as a classical field, whereas 
WIMPs are highly non-degenerate.  WIMPs have de Broglie 
wavelength of order 
$10^{-10}~{\rm cm} \left({{\rm GeV} \over m_W}\right)$
where $m_W$ is the WIMP mass ($\hbar = c = 1$).  A WIMP 
that misses a detector cannot deposit any energy into it.  
Furthermore, when a WIMP is detected no precise information 
is obtained regarding the location and velocity of any 
other WIMP.  Axions are bosons and much lighter than 
WIMPs.  A typical expectation for the mass of dark 
matter axions is $m_a \sim 10^{-5}$ eV.  In some 
models they are lighter than this by several orders 
of magnitude.  Their de Broglie wavelength is of order 
$\lambda \sim 100~{\rm m} \left({10^{-5}~{\rm eV} 
\over m_a}\right)$.  The number of axions in a volume 
of size $\lambda^3$ is of order $3 \times 10^{25}
\left({10^{-5}~{\rm eV} \over m_a}\right)^4$. The 
axion fluid surrounding us is therefore, in first
approximation, a classical field oscillating at 
the frequency 
\begin{equation}
\nu_a = 2.42~{\rm GHz}~\left({m_a \over 10^{-5}~{\rm eV}}\right)
\label{freq}
\end{equation}
with frequency dispersion $\delta \nu_a \sim 10^{-6} \nu_a$
due to the spread in kinetic energy.  The axion field has 
a coherence time of order
\begin{equation}
t_{\rm coh} \sim {1 \over \delta \nu_a} 
\sim 0.4~{\rm ms} \left({10^{-5}~{\rm eV} \over m_a}\right)~~\ .
\label{tcoh}
\end{equation}
The coherence time of individual caustic forming cold 
flows is much longer.

Once axion dark matter has been detected, the frequency 
of the axion field oscillations is known.  We also 
know the phase of the axion field at all times provided 
we keep measuring it on time scales short compared to 
$t_{\rm coh}$.  This should be achievable since the 
existing ADMX detector \cite{ADMX}, on resonance, 
converts on the order of thousand axions to photons 
per second.  If the cavity is driven with power and 
the oscillation frequency of the cavity is adjusted 
in time so that the phase of the axion field stays 
90$^\circ$ ahead of the phase of the oscillation 
stored in the cavity, a maximum amount of power is 
extracted from the axion field.  

When tuned to the axion mass, the cavity detector of dark 
matter axions behaves as a driven harmonic oscillator:
\begin{equation}
{d^2 X \over dt^2} + \gamma {dX \over dt} + 
\omega^2 X = f \cos(\omega t)~~\ .
\label{ho}
\end{equation} 
The quality factor of the oscillator is $Q = {\omega \over \gamma}$.
The steady state solution in case the cavity is driven only by the 
axion field is
$X(t) = {f \over \gamma \omega} \sin(\omega t)$.  
The power deposited into the cavity is then
\begin{equation}
P = {Q \over 2 \omega} f^2 = g^2 \rho_a B^2 V C Q {1 \over m_a} 
\label{pow}
\end{equation}
where $g$ is the coupling of the axion to two photons, 
$\rho_a$ is the axion dark matter density at the detector, 
$V$ is the volume of the cavity, $B$ the strength of the 
magnetic field permeating the cavity, and $C$ a number
of order one describing how strongly the relevant cavity 
mode couples to the axion field.  The power is of order 
$10^{-22}$ Watt for the ADMX cavity detector which has 
a volume of order 125 liters, a magnetic field of order 
7.5 Tesla, and a quality factor of order $10^5$.  Higher 
quality factors, of order $10^{10}$, have been achieved 
with superconducting cavities.  Unfortunately, the need
for a magnetic field inside the cavity makes the use 
of superconducting cavities problematic.  If the cavity 
is driven with power $P_0$, with phase 90$^\circ$ behind
that of the driving force due to the axion field, the 
power transfered to the cavity from axion dark matter
is $\sqrt{P_0 P}$.  Thus, if one drives the cavity with 
10 MW of power on resonance, the axion field contributes 
an additional $0.3 \times 10^{-7}$ W.  Of course, no one 
would want to go through so much trouble to generate such 
a small amount of power!  

The difficulty with extracting power from axion dark 
matter is that the coupling $g$ and the other factors
on the right hand side of Eq.~(\ref{pow}), $B$, $V$
and $Q$, are too small.  However let us imagine that 
the problem of weak coupling to the axion field has 
been solved by some clever trick, and derive the 
maximum amount of power that can be extracted from 
the axion fluid when the coupling is arbitrarily 
large. We model the axion detector/power plant as 
a harmonic oscillator coupled to the axion field 
$\phi(\vec{x},t)$ 
\begin{eqnarray}
({d^2 \over dt^2} + \gamma {d \over dt} + \omega^2) X(t)
&=& h \phi(\vec{0},t)
\nonumber\\
(\partial_t^2 - \nabla^2 + m_a^2) \phi(\vec{x},t)
&=& h X(t) \delta^3(\vec{x})~~\ .
\label{couple}
\end{eqnarray}
The detector is located at $\vec{x} = \vec{0}$. $h$ is 
a coupling with dimension (time)$^{-{1 \over 2}}$, which we are 
imagining to be large for the sake of argument.  In zeroth 
order in an expansion in powers of $h$, the axion field is 
taken to be a monochromatic wave coming by the detector: 
\begin{equation}
\phi^{(0)}(\vec{x},t) = A \cos(\vec{k}\cdot\vec{x} - \omega t) 
\label{ax0}
\end{equation}
where $\omega = \sqrt{\vec{k}\cdot\vec{k} + m_a^2}$.  
There is no prejudice in assuming the wave to be monochromatic
if the phase of the actual wave is known at all times.  Let
us assume that the oscillator, tuned to the same frequency 
$\omega$, is oscillating with amplitude $X_0$ and phase 
$\delta$ relative to the local axion field:
\begin{equation}
X(t) = X_0 \cos(\omega t + \delta)~~\ .
\label{Xosc}
\end{equation}
The power transferred from the axion field to the oscillator 
is then
\begin{equation}
P_X =~ <h {dX \over dt} \phi(\vec{0},t)>~ =~
- {1 \over 2} \omega h X_0 A \sin(\delta)~~\ .
\label{powtran}
\end{equation}
$P_X$ is largest when the oscillator lags 90$^\circ$ behind 
the local axion field.  The oscillator is a source of axion waves:
\begin{equation}
\phi^{(1)}(\vec{x},t) = {h X_0 \over 4 \pi r} 
\cos(kr - \omega t - \delta)
\label{phi1}
\end{equation}
where $r = |\vec{x}|$.  For $r >> \lambda$, the axion field 
Poynting vector is
\begin{eqnarray}
\vec{\cal P}(\vec{x}) &=& < - \partial_t \phi \vec{\nabla} \phi >
~=~ < - \partial_t(\phi^{(0)} + \phi^{(1)}) 
\vec{\nabla}(\phi^{(0)} + \phi^{(1)})>
\nonumber\\
&=& {1 \over 2} \omega \vec{k} A^2 
+ {h X_0 A \over 8 \pi r} \omega k (\hat{k} + \hat{r})
\cos(\vec{k}\cdot\vec{x} - kr + \delta) 
+ {h^2 X_0^2 \omega k \over 32 \pi^2 r^2} \hat{r}
\label{Poyn} 
\end{eqnarray}
where $\hat{k}$ and $\hat{r}$ are unit vectors in the 
directions of $\vec{k}$ and $\vec{x}$ respectively.  The 
first term in Eq.~(\ref{Poyn}) is the energy flux density 
in the original axion wave, Eq.~(\ref{ax0}).  The last 
term is the energy flux density in the field radiated 
by the oscillator, Eq.~(\ref{phi1}).  The total power 
radiated is 
\begin{equation}
P^{(2)} = {h^2 X_0^2 \omega k \over 8 \pi}~~\ .
\label{radpow}
\end{equation}
The second term in Eq.~({\ref{Poyn}) is the interference 
term between the original axion wave and the radiated
wave.  The corresponding outgoing power within a solid 
angle $\Delta \Omega$ is 
\begin{eqnarray}
P^{(1)} &=& \lim_{r \rightarrow \infty} 
\int_{\Delta \Omega} d \Omega ~r^2 ~
\hat{r}\cdot \vec{\cal P}^{(1)}(\vec{x})
\nonumber\\
&=&~~~ 0~~~~~~~~~~~~~~~~~~~~~~~~~{\rm if}~\hat{k}~{\rm points}~
{\rm outside}~\Delta \Omega
\nonumber\\ 
&=&~ + {1 \over 2} \omega h X_0 A \sin(\delta)~~~~~~{\rm if}~
\hat{k}~{\rm points}~{\rm inside}~\Delta \Omega~~\ .
\label{pow1}
\end{eqnarray}
The interference term $P^{(1)}$ accounts for the power $P_X$ extracted 
by the oscillator from the axion field.  Eq.~(\ref{radpow}) implies
a limit on how much power can be obtained. Even if we find a way to 
increase $h$, we will want to keep $P^{(2)} < P_X$.  Otherwise we 
lose more power in radiation than we extract from the axion field.  
This requires $h X_0 < {4 \pi \over k} A$ and hence
\begin{equation}
P_X < {4 \pi \over k^2} {\cal P}^{(0)} = 
1.3 \cdot 10^4~{\rm W} 
\left({10^{-5}~{\rm eV} \over m_a}\right)^2~~\ .
\label{final}
\end{equation}
In the axion case, if we were able to increase the 
coupling $h$, we would be able to collect up to the 
energy flowing within a radius ${2 \over k}
= {\lambda \over \pi}$ of the detector/power plant, 
but no more than that.  There is no physics principle 
that limits the size of $h$.  Unfortunately we do not 
know at present how to increase $h$ much in practice.

Let us also consider dark matter constituted of axion-like 
particles (ALPs).   These are light spin zero bosons like 
axions \cite{Arias} but they do not solve the Strong CP 
Problem of the Standard Model of particle physics.  ALPs 
are much less constrained than axions.  In particular 
their coupling strengths are not necessarily proportional 
to their mass (as is the case for axions) and they may be 
very light.  Consider for example dark matter constituted of 
ALPs of mass $10^{-15}$ eV.  Eqs.~(\ref{freq}), (\ref{tcoh}) 
and (\ref{final}) apply, so their oscillation frequency is  
0.24 Hz, their coherence time is of order 4 months and the 
maximum amount of power that can extracted from them is 
1.3$\cdot 10^{12}$ TW, which is quite a lot more than needed.  
The remaining issue is the coupling strength.  We may consider 
the ALP coupling to two photons which gives rise to Eq.~(\ref{pow}).
Let us assume that a high quality resonator can be found that 
exploits that coupling in the ALP case and that $B^2 V C Q$ 
has the same value as achieved in ADMX.  The power is boosted 
by the factor $1/m_a$.  Although $g$ is unrelated to $m_a$, 
it is bounded by the CAST solar axion search \cite{CAST}.   
Assuming $g$ saturates the CAST bound, we would get 
$P \sim 10^{-2}$ W.    

\vskip 0.5cm 

Happy Birthday Frank, and many happy returns!

\begin{acknowledgments}

This work was supported in part by the U.S. Department of Energy under grant 
DE-SC0010296 at the University of Florida.

\end{acknowledgments}



\end{document}